\documentclass[a4paper, 12pt]{panl}
\usepackage[T2A]{fontenc}
\usepackage[utf8x]{inputenc}
\usepackage[main=english,russian]{babel}
\usepackage{cite}
\usepackage{wrapfig}
\usepackage{graphicx}
\usepackage{amssymb}
\usepackage{amsfonts}
\usepackage{amsmath}
\usepackage{longtable}
\usepackage{rotating}
\usepackage{lscape}
\usepackage{epsfig}
\usepackage{multirow}

\originalTeX
\begin{document}

\newcommand{\wtb}{\textit{Wtb~}}
\newcommand{\twb}{\textit{tWb}}
\newcommand{\gm}{\gamma^\mu}
\newcommand{\smn}{\sigma^{\mu \nu}}
\newcommand{\Wmn}{W_{\mu \nu}}
\newcommand{\vl}{V_L}
\newcommand{\vr}{V_R}
\newcommand{\gl}{g_L}
\newcommand{\gr}{g_R}
\newcommand{\DMD}{\overleftrightarrow{D^\mu}}

\title{Separation of left-handed and anomalous right-handed vector operators contributions into the Wtb vertex for single and double resonant top quark production processes using a neural network}
\maketitle

\authors{E.\,Abasov $^{a}$\footnote{E-mail: emil@abasov.ru},
E.\,Boos$^{a}$, V.\,Bunichev$^{a}$, L.\,Dudko$^{a}$, }
\authors{D.\,Gorin$^{a}$, A.\,Markina$^{a}$, M.\,Perfilov$^{a}$, O.\,Vasilevskii$^{a}$,}
\authors{P.\,Volkov$^{a}$, G.\,Vorotnikov$^{a}$,A.\,Zaborenko$^{a}$} 
\vspace*{6pt}

\from{$^{a}$\,Skobeltsyn Institute of Nuclear Physics of Lomonosov Moscow State University (SINP MSU), 1(2), Leninskie gory, GSP-1, Moscow 119991, Russian Federation }

\setcounter{footnote}{0}
\vspace*{6pt}

\begin{abstract}
The paper describes the application of deep neural networks for the search of deviations from the Standard Model predictions at the \wtb vertex in the processes of single and double resonant top quark production with identical final state \twb. Monte-Carlo events preliminary classified by first level neural network as corresponding  to single or double resonant top quark production are analyzed by two second level neural networks if there is a possible contribution of the anomalous right-handed vector operator into \wtb vertex or events are corresponded to the Standard Model. The second level neural networks are different for single and double resonant classes. The classes depend differently on anomalous contribution and such splitting leads to better sensitivity. The developed statistical model is used to set constraints on the anomalous right-handed vector operator at the \wtb vertex in different regions of phase space. It is demonstrated that the proposed method allows to increase the efficiency of a search for the anomalous contributions to the \wtb vertex.

\end{abstract}
\vspace*{6pt}

\section*{Introduction}
\label{sec:intro}
One of the most important tasks of modern high-energy physics is a search for deviations from the predictions of the Standard Model (SM). The motivation for such research comes from the knowledge that the SM cannot be recognized as a final theory e.g. because of the presence of a significant number of free parameters or the fact of Dark Matter existence. 
The top quark, as the elementary particle with the largest mass in the SM is a object of close attention for study since there are substantial reasons to believe that physics beyond the SM can manifest itself in top quark sector first~\cite{Boos:2019ffk}. In particular, the element of the \textit{CKM} matrix $V_{tb}$ is closer to 1 than the other diagonal elements~\cite{ParticleDataGroup:2022pth}, which lacks a clear explanation. The \wtb vertex, which describes the interaction of the top quark with the \textit{W} boson and the \textit{b} quark in the SM, has a vector-axial structure, and any deviations from such structure would indicate to new physics beyond the SM. Therefore, it is important to investigate the structure of the \wtb vertex both phenomenologically and experimentally. There are experimental constraints on the parameters of anomalous interactions at the \wtb vertex obtained for \textit{t}-channel single top quark production processes~\cite{CMS:2016uzc} and for \textit{tW}-associated single top quark production processes~\cite{CMS:2019zct}. In the latter case, a scheme based on the removal of part of the diagrams was used to distinguish single resonant top quark production processes from double resonant production. 
It is demonstrated \cite{Baskakov:2017thj} that different schemes for separating single resonant top quark production processes from double resonant production have varying sensitivities to anomalous contributions at the \wtb vertex. It should not be forgotten that double resonant production top quark production processes, whose cross section significantly exceeds that of single top quark production processes, are also sensitive to anomalous contributions at the \wtb vertex, albeit to a lesser extent than single resonant top quark production processes because the \wtb vertex appears twice, in both the production and decay of top quarks, in the latter case. In \cite{Boos:2023kpp}, it is proposed to use a complete gauge-invariant set of diagrams corresponding to single and double resonant top production processes for analysis, without removing any part of it, and to separate these processes in phase space using deep neural networks. Subsequently, the selected events, classified by the neural network at the first level as events that are correspond to single or double resonant top production processes, can be used to analyze the possible presence of anomalous operators at the \wtb vertex in two separate classes, which depend differently on the anomalous operators.

This work is dedicated to search for anomalous contributions at the \wtb vertex applying neural networks. The paper is organized as follows. Section~\ref{sec:twb} briefly describes the method for separating single and double resonant top quark production processes with identical final states using neural networks. Section~\ref{sec:anom} compares two phenomenological approaches currently used to study anomalous contributions at the \wtb vertex. Section~\ref{sec:nn} provides a description of the neural networks that separate the contributions of different \wtb operators into a \wtb vertex and presents the results. Section~\ref{sec:stat} describes the statistical model used to obtain upper limits on the coupling parameter of the anomalous right-handed vector \wtb operator. The Conclusion compares the efficiencies of separating anomalous contributions at the \wtb vertex for different regions of phase space and outlines the prospects for further research on this topic.

The \textit{CompHEP} computational package~\cite{CompHEP:2004qpa, Pukhov:1999gg} is used for Monte Carlo event generation, numerical calculations and distributions construction. The \textit{Tensorflow} software package~\cite{Abadi:2016kic} is used for creation of the deep neural networks employed in this study.

\newpage
\section{The processes of single and double resonance production of the top quark with identical final states \twb}
\label{sec:twb}
The gauge invariant set of diagrams for the leading subprocess $\rm g,g\to l, \nu, b, \bar b, q, \bar q^\prime$ of the full process $\rm p,p\to l, \nu, b, \bar b, q, \bar q^\prime$ with \twb~final state is presented in~\cite{Boos:2023kpp}. This set of diagrams contains the diagrams that correspond to both single and double resonant top quark production. The separation of events corresponding to single and double resonant top quark production is performed using a neural network. In this case, the existing interference between the processes is "smeared" among the events classified by the neural network as belonging to one class or another.

It is important to note that contribution of the double resonant top quark production processes into the total cross section is dominate, however the dependence on the anomalous contributions into the \wtb vertex is weaker than that one of the single resonant top quark production processes which contribution to the total cross section is much smaller.

After successfully separating these processes in phase space using the neural network, the task of isolating anomalous contributions at the \wtb vertex can be undertaken separately for single and double resonant top quark production processes.

\section{The structure of the \wtb vertex with the presence of anomalous interaction parameters}
\label{sec:anom}

New physics beyond the electroweak scale can be parameterized by the Lagrangian of the following form:
\begin{equation}
\mathcal{L}^\mathrm{eff} = \sum \frac{C_x}{\Lambda^2} O_x + \dots \,,
\label{ec:effL}
\end{equation}
where $O_x$ are dimension-6 operators, invariant under the gauge symmetry of the Standard Model (SM), $\Lambda$ is the scale of new physics, and the dots represent higher-dimensional operators that are usually ignored. The classification of dimension-6 operators which ones can be used to parameterize contributions of new physics up to higher orders can be found in~\cite{Buchmuller:1985jz}, where the basis operators are also provided. Subsequently, it was shown in~\cite{Grzadkowski:2003tf},~\cite{Aguilar-Saavedra:2009ygx} that many of these operators can be discarded without loss of generality, simplifying the phenomenology. The most general form of the \wtb vertex Lagrangian can be written as~\cite{Aguilar-Saavedra:2010kfy}:
\begin{equation}
\mathfrak{L}=-\frac{g}{\sqrt{2}}\bar{b}\gamma^{\mu}\left( f_{\rm V}^{\rm L}P_{\rm L} + f_{\rm V}^{\rm R}P_{\rm R}\right)tW_{\mu}^{-} - \frac{g}{\sqrt{2}}\bar{b}\frac{i\sigma^{\mu\nu} \partial_{\nu} W_{\mu}^{-}}{M_{\rm W}}\left( f_{\rm T}^{\rm L}P_{\rm L} + f_{\rm T}^{\rm R}P_{\rm R}\right)t + h.c.
\label{lagrangian_wtb}
\end{equation}
where $P_{\rm L,R}=\frac{1\mp\gamma_{5}}{2}$, $\sigma_{\mu\nu}=\frac{i}{2}(\gamma_{\mu}\gamma_{\nu} -\gamma_{\nu}\gamma_{\mu})$, the form factor $f_{\rm V}^{\rm L}$ ($f_{\rm V}^{\rm R}$) is a parameter that determines the magnitude of the left-handed (right-handed) vector operator, and $f_{\rm T}^{\rm L}$ ($f_{\rm T}^{\rm R}$) is a parameter that determines the magnitude of the left-handed (right-handed) tensor operator.

Here $f_{\rm V}^{\rm L} = V_{tb} + C_{\phi q}^{(3,3+3)*} \frac{v^2}{\Lambda^2}$, $f_{\rm V}^{\rm R} = \frac{1}{2} C_{\phi \phi}^{33*} \frac{v^2}{\Lambda^2}$, $f_{\rm T}^{\rm L} = \sqrt 2 C_{dW}^{33*} \frac{v^2}{\Lambda^2}$, $ f_{\rm T}^{\rm R} = \sqrt 2 C_{uW}^{33} \frac{v^2}{\Lambda^2}$ are gauge invariant operators.

\begin{align}
& O_{\phi q}^{(3,3+3)} = \frac{i}{2} \, \left[ \phi^\dagger (\tau^I D_\mu
  - \overleftarrow D_\mu \tau^I)  \phi \right] (\bar q_{L3} \gm \tau^I q_{L3}) \,,
&& {O_{\phi \phi}^{33}} = i (\tilde \phi^\dagger D_\mu \phi)
        (\bar t_{R} \gm b_{R}) \,, \notag \\
& O_{dW}^{33} = (\bar q_{L3} \smn \tau^I b_{R}) \phi \, \Wmn^I \,,
&& O_{uW}^{33} = (\bar q_{L3} \smn \tau^I t_{R}) \tilde \phi \, \Wmn^I \,,
\end{align}

In the SM, the form factors take the following values: $f_{\rm V}^{\rm L} = V_{tb};  f_{\rm V}^{\rm R} = f_{\rm T}^{\rm L} = f_{\rm T}^{\rm R} = 0$.

In phenomenological analyses, deviations from the SM are usually sought; a standard approach is to consider scenarios where the left-handed vector operator with parameter $f_{\rm V}^{\rm L}$ (SM-like structure) and one of the anomalous operators are presented in (\ref{lagrangian_wtb}). The $(L_{\rm V}, R_{\rm V})$ scenario where, in addition to the left-handed vector operator present in the SM, a right-handed anomalous vector operator is also present at the \wtb vertex is considered here. This scenario is most difficult for distinguish the deviations from SM prediction.

\section{Separation of various operators contributions at the \wtb vertex using deep neural networks}
\label{sec:nn}

In~\cite{Boos:2023kpp}, a method for separating single and double resonant top quark production processes in phase space using deep neural networks is described and results are presented. This section describes how events, pre-selected by a neural network and splitted as corresponding to a specific process, can be subsequently classified by other neural networks depending on the contribution of the left-handed or anomalous right-handed vector operator at the \wtb vertex in these processes.

For training the deep neural networks designed to separate the contributions of anomalous operators in the \(\wtb\) vertex for the \((L_{\rm V}, R_{\rm V})\) scenario, it was necessary to simulate sets of Monte-Carlo (MC) events corresponding to two subsets of diagrams from the complete set, as well as different values of anomalous coupling parameters in the \(\wtb\) vertex from (\ref{lagrangian_wtb}) for the \((L_{\rm V}, R_{\rm V})\) scenario.

The created event sets correspond to the full set of diagrams with the final state \(\twb\) (with subsequent decays of the top quark and \(W\) boson), containing diagrams of both single and double resonant top quark production (denoted here and further as "\(\twb\)"); a subset of diagrams containing single resonant top quark production diagrams ("DR") and a subset of diagrams containing double resonant top quark production diagrams ("\(t\bar{t}\)") for the values of anomalous coupling parameters \(f_{\rm V}^{\rm L} = 1, f_{\rm V}^{\rm R } = 0\) ("\textit{SM}" case) and \(f_{\rm V}^{\rm L} = 0, f_{\rm V}^{\rm R } = 1\) ("\textit{RV}").

\begin{figure}[ht]
\begin{minipage}[b]{.49\linewidth}
\centering
\includegraphics[width=.95\linewidth]{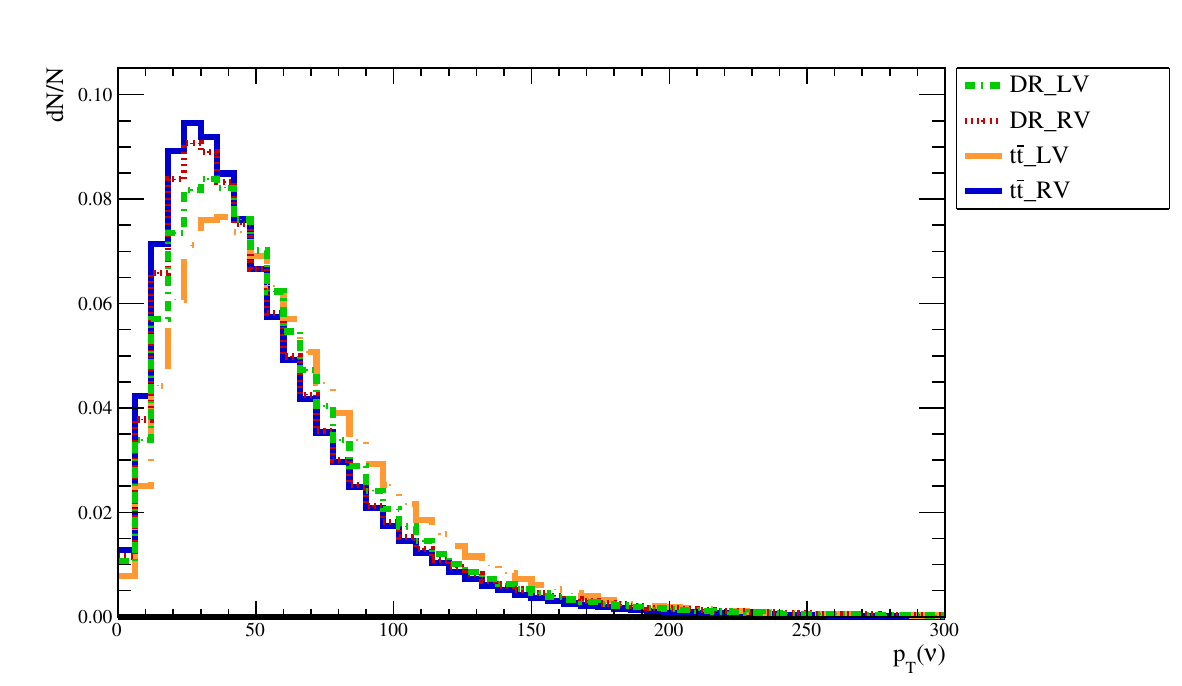}
\includegraphics[width=.95\linewidth]{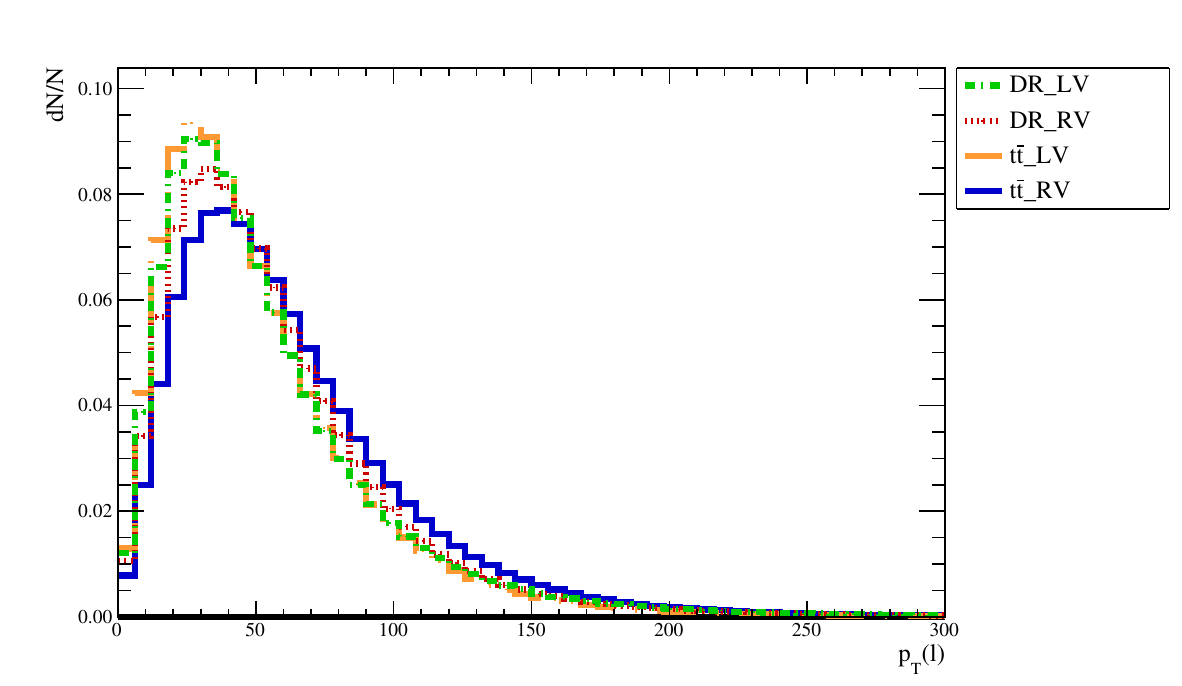}
\end{minipage}
\begin{minipage}[b]{.49\linewidth}
\centering
\includegraphics[width=.95\linewidth]{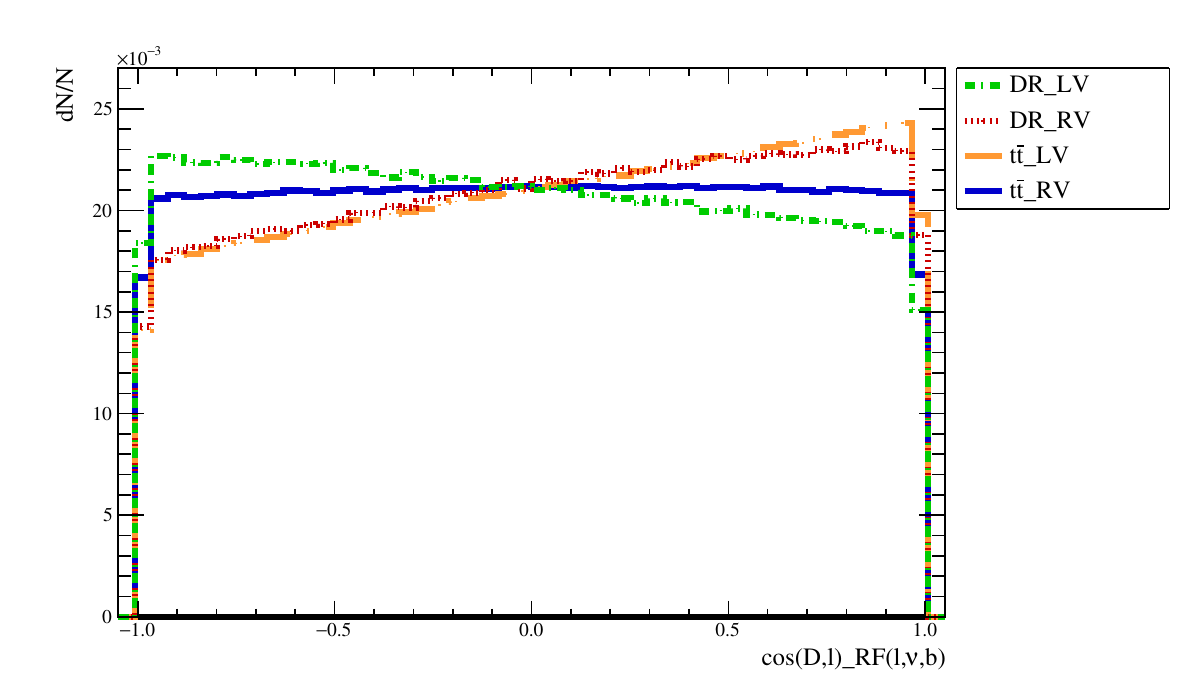}
\includegraphics[width=.95\linewidth]{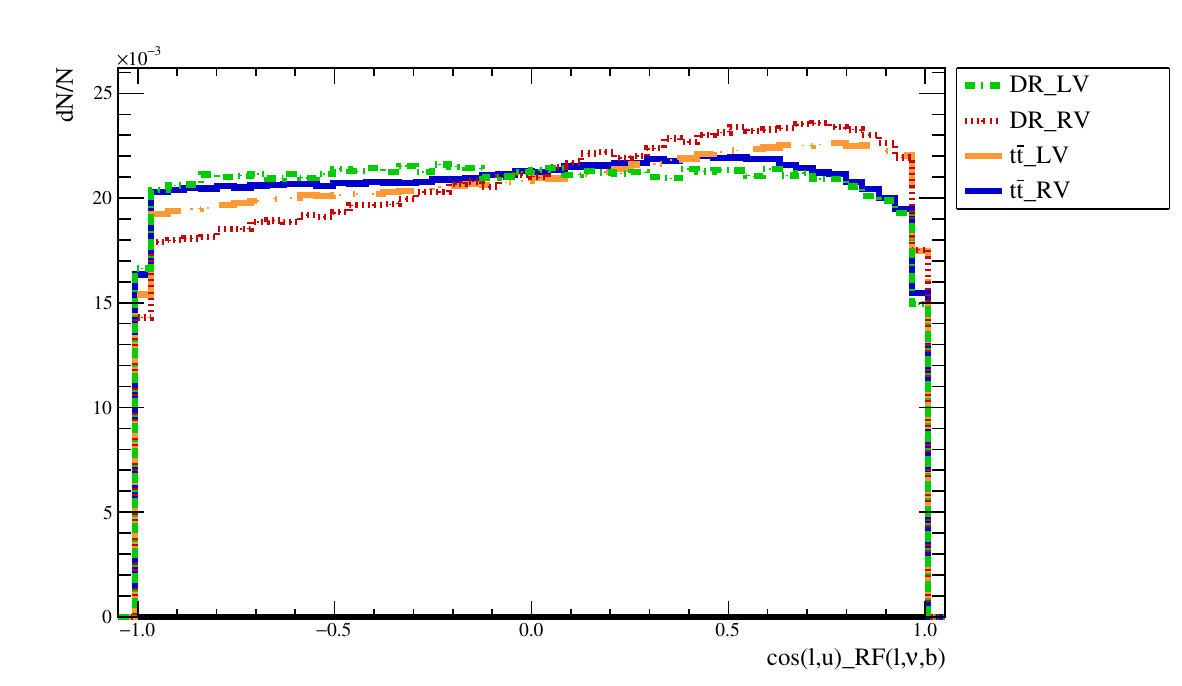}
\end{minipage}
\caption{The distribution of simulated MC events across representative kinematic variables used for the training of neural network to distinguish contribution of different operators in the \(\wtb\) vertex for selected phase space regions corresponding to single and double resonant top quark production processes.} \label{fig:nn_vars}
\end{figure}

It is also necessary to select kinematic variables whose behavior differs significantly depending on which operator presents in the \(\wtb\) vertex. To determine such variables, previously developed universal methods for forming high-level observables (a set of optimal observables based on the analysis of Feynman diagrams for a specific task) and basic low-level observables (a universal set of basic observables that does not account for the specifics of the task but allows deep neural networks to identify desired patterns) were applied~\cite{Boos:2008sdz},~\cite{Abasov:2022byr}. The distributions of simulated MC events across some kinematic variables used in training the neural network to distinguish events corresponding to different contributions of anomalous operators in the \(\wtb\) vertex for selected phase space regions corresponding to single  and double resonant top quark production processes are shown in Fig.~\ref{fig:nn_vars}. Four curves in each plot are obtained from the created MC event sets corresponding to single or double resonant top quark production in the presence of only left-handed ("SM") or only right-handed ("RV") vector operators. It is evident that the distributions for different processes and different operators in the \(\wtb\) vertex differ, allowing the use of these variables (along with many others) for training the neural network that separates contributions from different \(\wtb\) operators in phase space regions where single  or double resonant top quark production processes dominate.

\begin{figure*}
\begin{minipage}[b]{.49\linewidth}
\centering
\includegraphics[width=.95\linewidth]{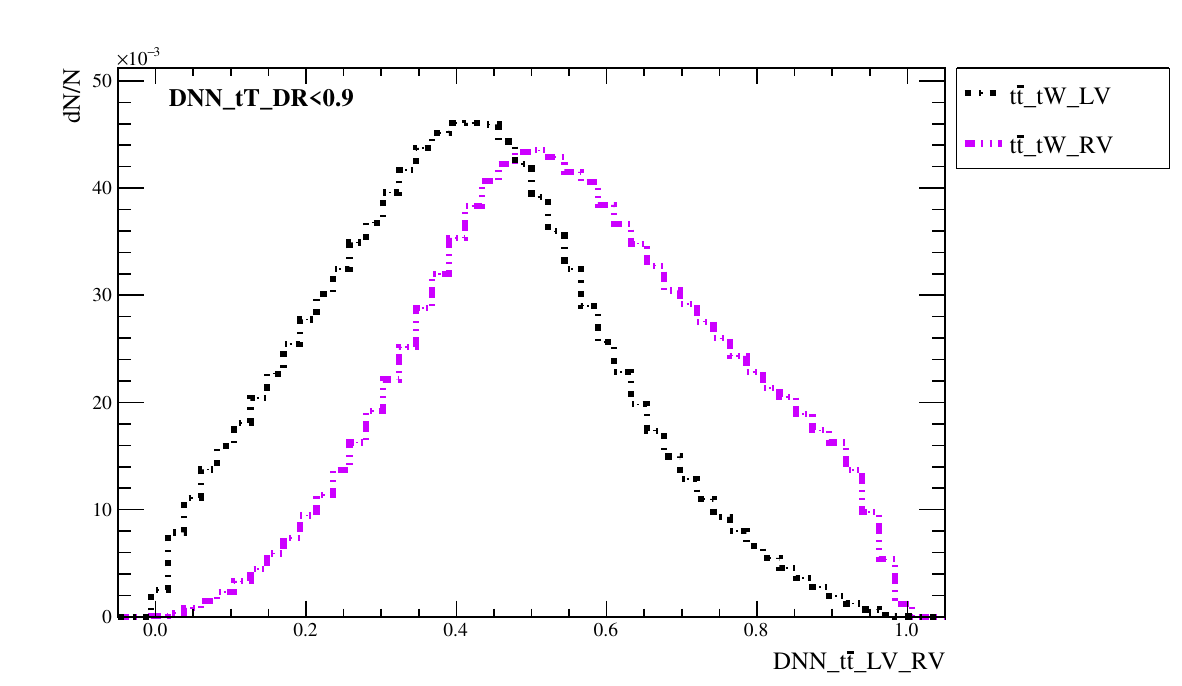}
\caption{The distribution of simulated events by the values of the neural network discriminator, distinguishing contributions from the left-handed and anomalous right-handed vector operators in the \(\wtb\) vertex for events classified by the previous neural network as double resonant top quark production events.} \label{fig:nn_tT}
\end{minipage}
\begin{minipage}[b]{.49\linewidth}
\centering
\includegraphics[width=.95\linewidth]{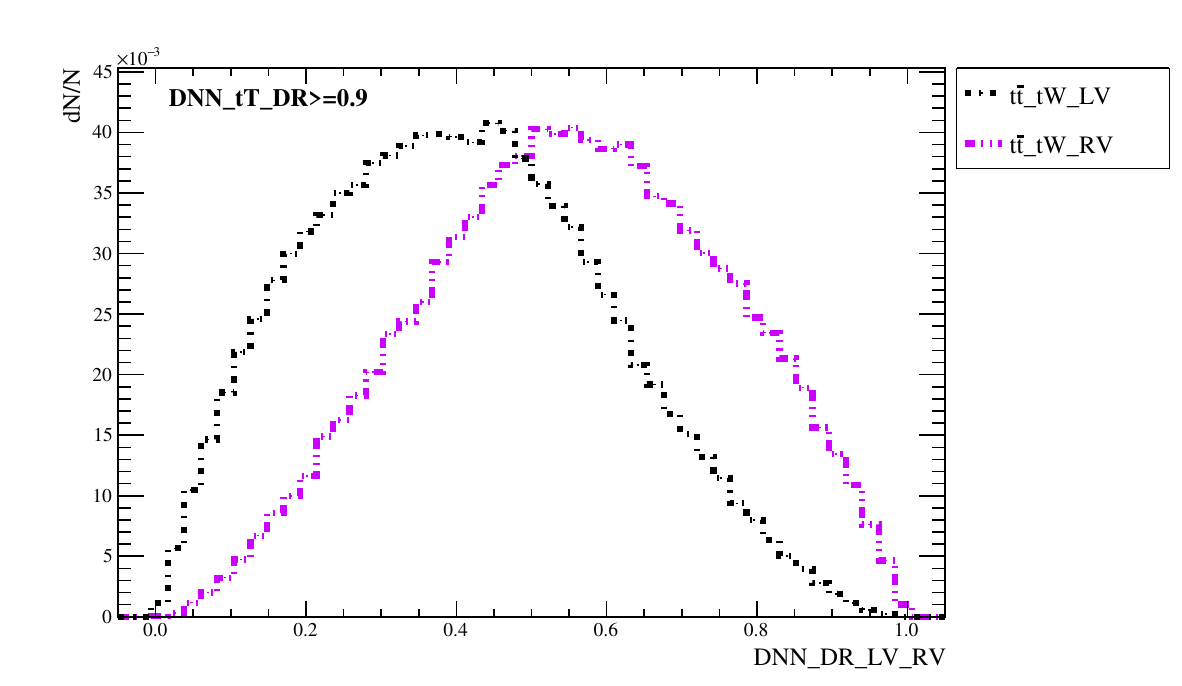}
\caption{The distribution of simulated events by the values of the neural network discriminator, distinguishing contributions from the left-handed and anomalous right-handed vector operators in the \(\wtb\) vertex for events classified by the previous neural network as single resonant top quark production events.} \label{fig:nn_tWb}
\end{minipage}
\end{figure*}

Figure~\ref{fig:nn_tT} shows the distribution of simulated events by the values of the neural network discriminator, distinguishing contributions from the left-handed and right-handed anomalous vector operators in the \(\wtb\) vertex for events classified by the previous neural network as double resonant top quark production events (the neural network discriminator value assigned to the selected events at the previous stage is less than 0.9). Similar separation is demonstrated in Fig.~\ref{fig:nn_tWb}, but for events classified by the previous neural network as single resonant top quark production events.

It can be seen that in both cases good separation is achieved. The next section describes a statistical model that allows, based on the data of events classified by neural networks as having different operator contributions in the \(\wtb\) vertex in phase space regions where single or double resonant top quark production processes prevail, to obtain upper limits on the coupling parameter \(f_{\rm V}^{\rm R}\) of the anomalous right-handed vector operator in the \(\wtb\) vertex.

\section{Statistical model to constrain the anomalous interaction parameters in the \wtb vertex}
\label{sec:stat}
A full-fledged statistical analysis involves a detailed estimation of various systematic uncertainties, forms a statistical model of all nuisances and find probability density function for the parameter of interest. For the simplified phenomenological analysis an approximate estimation of the upper limits on \(\left\vert f_{\rm V}^{\rm R} \right\vert\) can be done with asymptotic formulas  based on~\cite{Basso:2021dwp,Cowan:2010js,gorin2024asimptoticheskie15008} and used to compare sensitivity of different approaches. 

We briefly outline this derivation. Assuming that the data follows a Poisson distribution in two regions: the signal (RV) and the control (SM) regions, with histograms having N and M bins respectively, the logarithm of the likelihood function is given by:
\begin{equation}
\begin{aligned}
\ln L(\mu, B)=\sum_{i=1}^N\left(n_i \cdot \ln \left(\frac{\mu s_i + b_i}{n_i!}\right)-\mu s_i - b_i\right) + \\
+ \sum_{j=1}^M\left(m_j \cdot \ln \left(\frac{\tau b_j}{m_j!}\right) - \tau b_j\right) \label{eq:4}
\end{aligned}
\end{equation}
Here, \(n_i\) and \(m_i\) are the number of events in the i-th bin of the signal and control histograms, respectively, \(s_i, b_{i}\) are the number of signal and background events in the i-th bin, \(\mu\) is the signal strength, \(\tau\) is the background strength parameter, accounting for systematic uncertainties.

The systematic uncertainty of the integrated cross-section \(\Delta\) is assumed to be 20\%, corresponding to a more conservative error value used in real experimental analyses. To describe it, the parameter \(\delta_i = \frac{1}{\sqrt{\tau b_i}}\) is introduced, where \(\tau = \Delta^2 * \sum_{i=1}^M b_i\).

Next, we need to find the values of the parameters \(\hat{\hat{b}}_i\) that maximize the likelihood function for a given \(\mu\):
\begin{equation}
\begin{aligned}
\frac{\partial \ln L(\mu, \hat{\hat{b}}_i)}{\partial \hat{\hat{b}}_i} =  \left(\frac{n_i}{\mu s_i + \hat{\hat{b}}_i} - 1\right) + \\
+ \left(\frac{m_j}{\tau \cdot \hat{\hat{b}}_i} - \tau\right) = 0 , \\ \label{eq:5}
\end{aligned}
\end{equation}

For brevity, defining:
\begin{equation}
    S_1 = \sum_{i=1}^N\left(n_i \cdot \ln \left(\frac{\mu s_i + \hat{\hat{b_{ i}}}}{n_i}\right) + n_i - \mu s_i - \hat{\hat{b_{i}}}\right)\\ \label{eq:6}
\end{equation}
\begin{equation}
    S_2 = \sum_{j=1}^M\left(m_j \cdot \ln \left(\frac{\ \hat{\hat{b}}_j}{b_j}\right) + \tau \cdot (b_j - \hat{\hat{b}}_j)\right) \label{eq:7}
\end{equation}
        
We get for the significance \(Z\) in the Wilks \cite{Wilks:1938dza} and Wald \cite{Wald1943TestsOS} approximations, using formulas \eqref{eq:6} and \eqref{eq:7}:
\begin{equation}
\begin{aligned}
Z = \sqrt{-2\ln \frac{L(\mu, \hat{\hat{B}})}{L(\mu, B)}} = \sqrt{2 \left(S_1 + S_2\right)} \label{eq:Z}
\end{aligned}
\end{equation}
Thus, to find \(\mu\), it is necessary to solve the system of equations \ref{eq:5} and \ref{eq:Z}, substituting \(\tau = \frac{1}{\delta_i^2 \cdot b_i}\)\ and assuming Asimov data, which for the exclusion case \cite{Cowan:2010js} yields values: \(n_i = b_i\), \(m_i = \delta_i^{-2}\), setting \(Z = 1.65\) for a one-sided constraint with a 95\% confidence level.

Using these formulas, calculations were performed for two cases: in the whole phase space region  \(t\Bar{t} + tWb\), with \(\mu = {f_{\rm V}^{\rm R}}^2\), and with the splitting of phase space to two regions defined by first level neural network, where for the \(t\Bar{t}\) region the dependence \(\mu = {f_{\rm V}^{\rm R}}^2\), and for \(tWb\) - \(\mu = {f_{\rm V}^{\rm R}}^4\), based on the analysis of the general dependence of the numerator and denominator of the matrix element of single and pair top quark production processes on the coupling parameter with the anomalous right-handed vector operator~\cite{MohammadiNajafabadi:2006tfl}. The numerical solution of the equations gives the constraints \(\left\vert f_{\rm V}^{\rm R} \right\vert < 0.21\) without the proposed splitting and \(\left\vert f_{\rm V}^{\rm R} \right\vert < 0.17\) with the splitting of the phase space to single and double resonant regions.

\section*{Conclusion}
\label{sec:outro}

In the paper a method for separating the contributions of the left-handed and anomalous right-handed vector operators for processes of double resonant and single resonant top quark production using neural network is presented. Events with the final state \(\twb\) (with subsequent decays of the top quark and the \(W\)-boson) are initially separated in phase space using a deep neural network. The MC events, classified by the first level neural network as corresponding to single or double resonant top quark production, are subsequently analyzed by second level neural networks trained to distinguish the contributions of the anomalous right-handed vector operator from the left-handed vector operator in the \(\wtb\) vertex, in single and double resonant phase space regions separately. 

Using the developed statistical model, constraints were obtained on the coupling parameter of the right-handed vector operator in the \(\wtb\) vertex are obtained. These constraints are obtained separately for events with the final state \(\twb\), which were not divided into those corresponding to single or double resonant top quark production, and for events splitted by the neural network as corresponding to single or double resonant contributions. 

The constraints obtained by splitting the phase space regions corresponding to single or double resonant top quark production processes are stricter than when considering the undivided process with the final state \(\twb\). This improvement enhances the effectiveness of distinguishing the contributions of anomalous operators in the \(\wtb\) vertex using the proposed methods. The proposed approach gives possible to take into account the interference between single and double resonant top quark production in tWb final state with contribution of anomalous operators correctly. The approach minimizes systematic uncertainty and increases the sensitivity of the search for anomalous contributions.

\section*{Acknowledgements}
\label{sec:acknowledgement}
This work is conducted with financial support of grant RSCF 22-12-00152.



\end{document}